\documentclass[12pt]{article}

\usepackage{graphicx}
\usepackage{epsfig}

\begin{document}


\begin{center}
{\bf\Large Hadroproduction of $\varphi$-mesons in \\the Quark-Gluon String model} \\ 
\vspace{1.cm}

{\bf G.H. Arakelyan$^1$, C. Merino$^2$, and Yu.M. Shabelski$^{3}$} \\

\vspace{.7cm}
$^1$A.Alikhanyan National Scientific Laboratory \\
(Yerevan Physics Institut)\\
Yerevan, 0036, Armenia\\
e-mail: argev@mail.yerphi.am\\
\vspace{0.1cm}

$^2$Departamento de F\'\i sica de Part\'\i culas, Facultade de F\'\i sica\\
and Instituto Galego de F\'\i sica de Altas Enerx\'\i as (IGFAE)\\
Universidade de Santiago de Compostela
Santiago de Compostela 15782 \\
Galiza-Spain\\
e-mail: merino@fpaxp1.usc.es\\
\vspace{0.1cm}

$^{3}$Petersburg Nuclear Physics Institute\\
NCR Kurchatov Institute\\
Gatchina, St.Petersburg 188350, Russia\\
e-mail: shabelsk@thd.pnpi.spb.ru
\vskip 0.9 truecm

\vspace{1.2cm}

{\bf Abstract}
\end{center}

We consider the experimental data on $\varphi$-meson production in 
hadron-nucleon collisions for a wide energy region. The Quark-Gluon String
Model quantitatively describes the spectra of secondary $\varphi$, as well as 
the ratios of $\varphi$/$\pi^-$ and $\varphi$/$K^-$ production cross 
sections.

\vskip 1.5cm


\newpage

%
\vskip -1.5cm
\section{Introduction}

The Quark-Gluon String Model (QGSM)~\cite{KTM,K20}, based on the Dual
Topological Unitarization (DTU), Regge phenomenology, and nonperturbative
notions of QCD, has been foe already more than thirty years
to to succesfully predict and describe many features of the hadronic processes
in a wide energy range. In particular, the QGSM allows one to make
quantitative predictions on the inclusive densities of different secondaries both in
the central and beam fragmentation regions. 

In the QGSM frame, high energy
hadron-nucleon collisions are considered as taking place via the exchange
of one or several Pomerons. Each Pomeron is considered in DTU as a cylindrical
diagram (Fig.~1a). The cut~\cite{AGK} between Pomerons in a multipomeron 
diagram results in elastic or diffraction dissociation processes, while the cut 
through one (Fig.~1b) or several (Fig.~1c) Pomerons corresponds to inelastic 
processes with multiple production of secondaries, the cut of every Pomeron
leading to the production of two showers of secondaries.
\vskip 2.cm
\begin{figure}[htb]
\label{cil}
\centering
\vskip -8.cm
\includegraphics[width=.7\hsize]{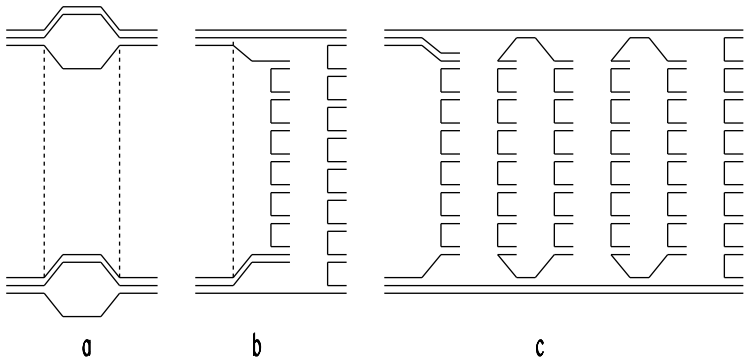}
\vskip -2.5cm
\caption{\footnotesize
(a) Cylindrical diagram representing a Pomeron exchange within the DTU
classification (quarks are shown by solid lines); (b) A cut of the 
cylindrical diagram corresponding to the single-Pomeron exchange contribution 
to inelastic $pp$ scattering; (c) The cuts of the diagrams for the inelastic 
interaction of the incident proton with a target nucleon in a $pp$ collision.}
\end{figure}

This model has been successfully used for the description of
multiparticle production processes in hadron-hadron 
collisions. The QGSM description of the production of secondaries (pseudoscalar mesons $\pi$
and $K$), and of nucleons $p$, $\overline{p}$, which give the main contribution to
mean multiplicity at different energies was obtained many years ago in~\cite{KaPi,Sh} (see
also~\cite{AMPS,MPS}). Vector meson production was 
considered in~\cite{aryer,yer,APSh}. The yields of hyperons, including the 
multistrange ones, has been described in~\cite{ACKS,Sigma}.

In the present paper, we apply first time the QGSM formalism to the description of the spectra 
of vector $\varphi$-mesons production in $\pi p$ and $pp$ collisioms, and of the ratios 
of yields $\varphi$/$\pi^-$ and $\varphi$/$K^-$ in $pp$-collisions for a large scope of the 
initial energy going up to the RHIC and LHC ranges. The $\varphi$-meson is a system,
of $s\overline{s}$ quarks with non-zero masses, that is rarely produced, and that it can be
thus sensitive to the production mechanism.

\section{Meson production in the QGSM}

In the QGSM the inclusive spectrum of a secondary hadron $h$ is
determined~\cite{KTM,K20} by the convolution of the diquark, valence quark, and sea 
quark distributions, $u(x,n)$, in the incident particles, with the 
fragmentation functions, $G^h(z)$, of quarks and diquarks into the secondary hadron $h$.
Both the distribution and the fragmentation functions are constructed using the Reggeon counting 
rules~\cite{Kai}.

For a nucleon target, the inclusive rapidity, $y$, or Feynman-$x$, $x_F$,
spectrum of a secondary hadron $h$ has the form~\cite{KTM}:
\begin{equation}
\frac{dn}{dy}\ = \
\frac{x_E}{\sigma_{inel}}\cdot \frac{d\sigma}{dx_F}\ = 
\sum_{n=1}^\infty w_n\cdot\phi_n^h (x) + w_D \cdot\phi_D^h (x) \ ,
\end{equation}
where the functions $\phi_{n}^{h}(x)$ determine the contribution of diagrams
with $n$ cut Pomerons, $w_n$ is the relative weight of this diagram, that
it depends on the beam particle, 
and the term $w_D \cdot\phi_D^h (x)$ accounts for the contribution of 
diffraction dissociation processes.

In the case of $pp$ collisions:
\begin{eqnarray}
\phi_n^{h}(x) &=& f_{qq}^{h}(x_{+},n) \cdot f_{q}^{h}(x_{-},n) +
f_{q}^{h}(x_{+},n) \cdot f_{qq}^{h}(x_{-},n)\\ 
&+& 2(n-1)\cdot f_{s}^{h}(x_{+},n) \cdot f_{s}^{h}(x_{-},n)\ \  , \nonumber
\end{eqnarray}
\begin{eqnarray}
x_{\pm} &=& \frac{1}{2}[\sqrt{4m_{T}^{2}/s+x^{2}}\pm{x}]\ \ , 
\end{eqnarray}
where $f_{qq}$, $f_{q}$, and $f_{s}$ correspond to the contributions
of diquarks, valence quarks, and sea quarks, respectively.

In the case of meson-nucleon collisions, the diquark contribution 
$f_{qq}^{h}(x_{+},n)$ in Eq.~(2) should be replaced by the valence antiquark
contributions $f_{\bar{q}}^{h}(x_{+},n)$. Thus, in the case of meson-nucleon
collisions, the quark and antiquark contributions
would be determined, respectively, by the convolution of the quark and
antiquark distributions with the corresponding fragmentation functions, e.g.
\begin{equation}
f_{q}^{h}(x_{+},n) = \int_{x_{+}}^{1}
u_{q}(x_{1},n)\cdot G_{q}^{h}(x_{+}/x_{1}) dx_{1}\ \ .
\end{equation}
The details of the model are presented in~\cite{KTM,K20,KaPi,Sh,ACKS}, and we
use in this paper the Pomeron parameters in ref.~\cite{Sh}.

The averaged number of exchanged Pomerons in $pp$ collisions
$\langle n \rangle_{pp}$ slowly increases with the energy.
In particular, in the case of $n > 1$, i.e. in the 
case of multipomeron exchange, the distributions of valence quarks and 
diquarks are softened due to the appearance of a sea quark contribution 
\cite{KTMS}. 

The production of $\pi$ and $K$ mesons in $pp$ collisions, starting from 
comparatively low energies, was analyzed in~\cite{KaPi,Sh}.

For the $\varphi$-meson production we use the following quark fragmentation
functions~\cite{aryer}:
\begin{eqnarray}
G_{u}^{\varphi} &=& G_{d}^{\varphi} =
a_{\varphi}\cdot(1-z)^{\lambda - \alpha_R - 2 \alpha_{\varphi}+2}, \;\; \\ 
G_{s}^{\varphi} &=& a_{\varphi}\cdot(1-z)^{\lambda - \alpha_{\varphi}}.
\end{eqnarray}
On the other hand, the diquark fragmentation functions into $\varphi$-mesons have the form:
\begin{equation}
G_{uu}^{\varphi} = G_{ud}^{\varphi} = 
a_{\varphi}\cdot(1-z)^{\lambda+\alpha_R -2(\alpha_R +\alpha_{\varphi})} \ \ ,
\end{equation}
where the parameter $\lambda$ takes the value $\lambda$=0.5~\cite{KTM,KaPi,Sh}, and the
parameters $\alpha_R$=0.5 and $\alpha_{\varphi}$=0. are the intercepts of the $\rho$
and $\varphi$ Regge trajectories, respectively.
The value of the parameter $a_{\varphi}$ is determined by comparison to experimental 
data on $\phi$ production from different hadron beams. In our calcualtions we
use the value $a_{\varphi}=0.11$.

\section{Numerical results}

In this section we compare the QGSM calculations to the experimental data on
$\varphi$ inclusive cross sections in $\pi p$~\cite{apsimon,agbenpi,daum}
and $pp$~\cite{daum,NA49,dijkstra,agben,brenner} collisions at different energies,
and to the experimental data on ratios of $\varphi/\pi$ and 
$\varphi/K$~\cite{starphi,alvec,alfik} for energies up to the LHC range. 

In Fig.~2 we compare two sets of experimental data for $x_F$-spectra of $\varphi$-mesons
produced in $\pi^{\pm} p$ collisions for two different initial momenta of $\pi$-mesons,
93$-$140 GeV/c (upper panel) and 175$-$360 GeV/c (lower panel), measured by different 
collaborations \cite{apsimon,agbenpi,daum,dijkstra}, to the corresponding QGSM calculations.
The full curve on the upper panel was calculated at 140 GeV/c, while the full curve on the
lower panel corresponds to momenta 200 GeV/c and the dashed curve to momenta 360 GeV/c.

As we can see on Fig.2, for $\pi p$-collision there is a remarkable disagreement between  
experimental data \cite{apsimon,daum} and \cite{dijkstra}, these being significantly higher.
The data \cite{apsimon,agbenpi,daum} were measured on proton target, while data \cite{dijkstra} 
were measured on Be target, and the absolute cross section per nucleon have been obtained using 
linear A-dependence. As it was mentioned in \cite{dijkstra}, the origin of the 
difference in the normalization has not been found. 

The theoretical QGSM curves for $\pi^+$ and $\pi^-$ beams practically coincide.

The QGSM description of the experimental data on the $x_F$ dependence of $d\sigma/dx_F$-spectra
of $\varphi$-mesons at $pp$-collision measured at different energies \cite{agbenpi,daum,NA49,dijkstra} 
is presented in Fig.~3, where one can see that the difference between the normalisation of different 
experiments in pp collisions is not so large as for $\pi p$ collisions.
The full curve in Fig.~3 corresponds to the QGSM calculations at 158 GeV/c.
Here, the agreement of the QGSM calculation with
the experimental data~\cite{NA49} is remarkably good. For the determination
of the normalisation parameter $a_{\varphi}$ experimental results by \cite{apsimon,agbenpi,NA49}
have been used. 

In Fig.~4, we compare the QGSM calculations to the experimental data on the inclusive spectra  
$x_F\cdot d\sigma /dx_F$ of $\varphi$-mesons in $pp$-collisions at 400 Gev/c~\cite{agben}.
In Fig.~5, the $y$- spectra $dn/dy$ of $\varphi$-mesons production in 
$pp$-collisions at 400 Gev/c~\cite{agben} are compared to the QGSM calculations at the same energy. 
The theoretical predictions for the same spectra at LHC energies 7 (dashed curve) 
and 14 Tev (dashed-dotted curve) are also presented.  

As it is shown in Figs.~2 to 5, the QGSM description of the experimental data at
intermediate energies is consistently satisfactory.

\begin{figure}[htb]
\label{pi140}
\centering
\vskip -8.cm
\includegraphics[width=.8\hsize]{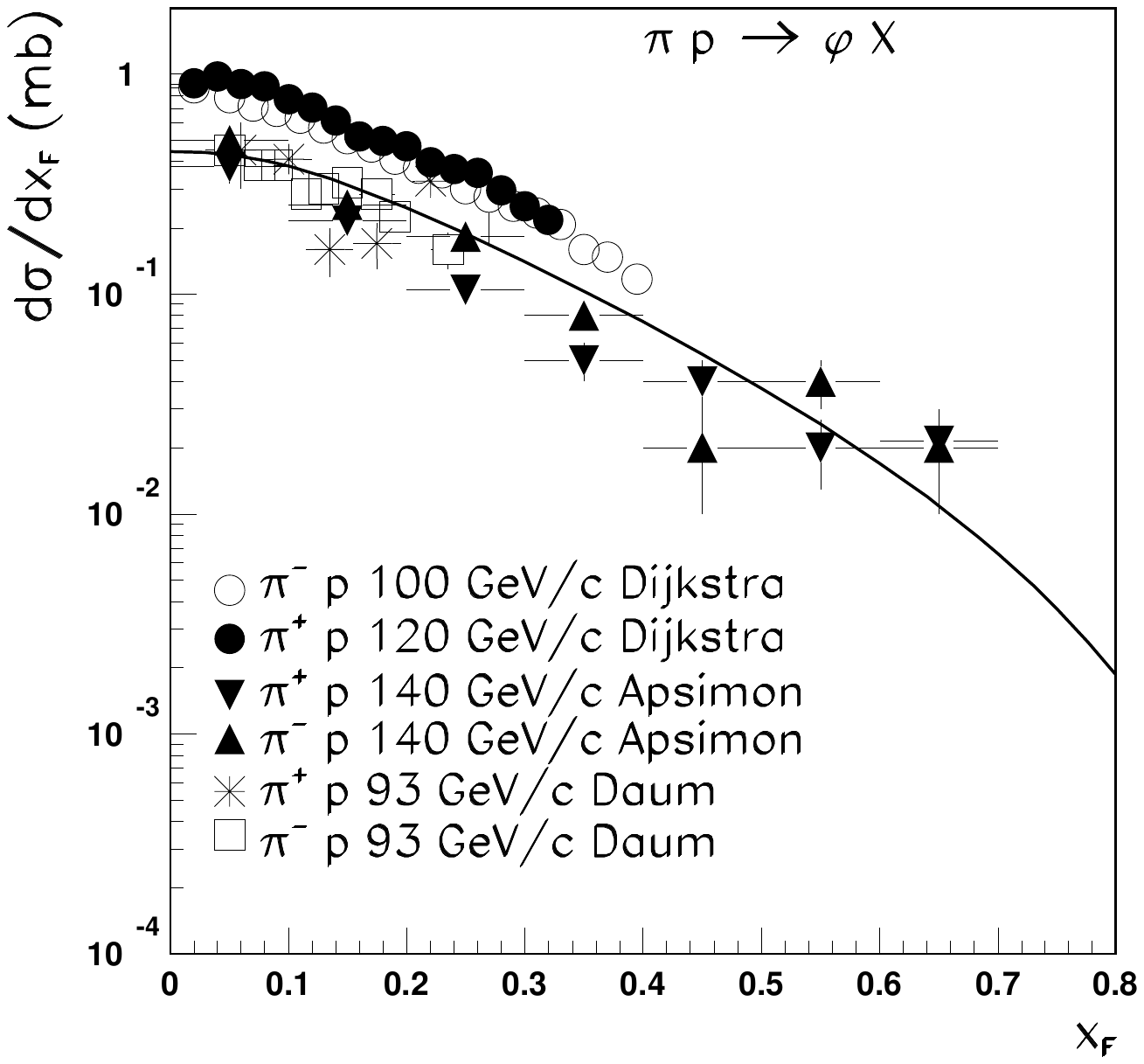}
\vskip -8.cm
\includegraphics[width=.8\hsize]{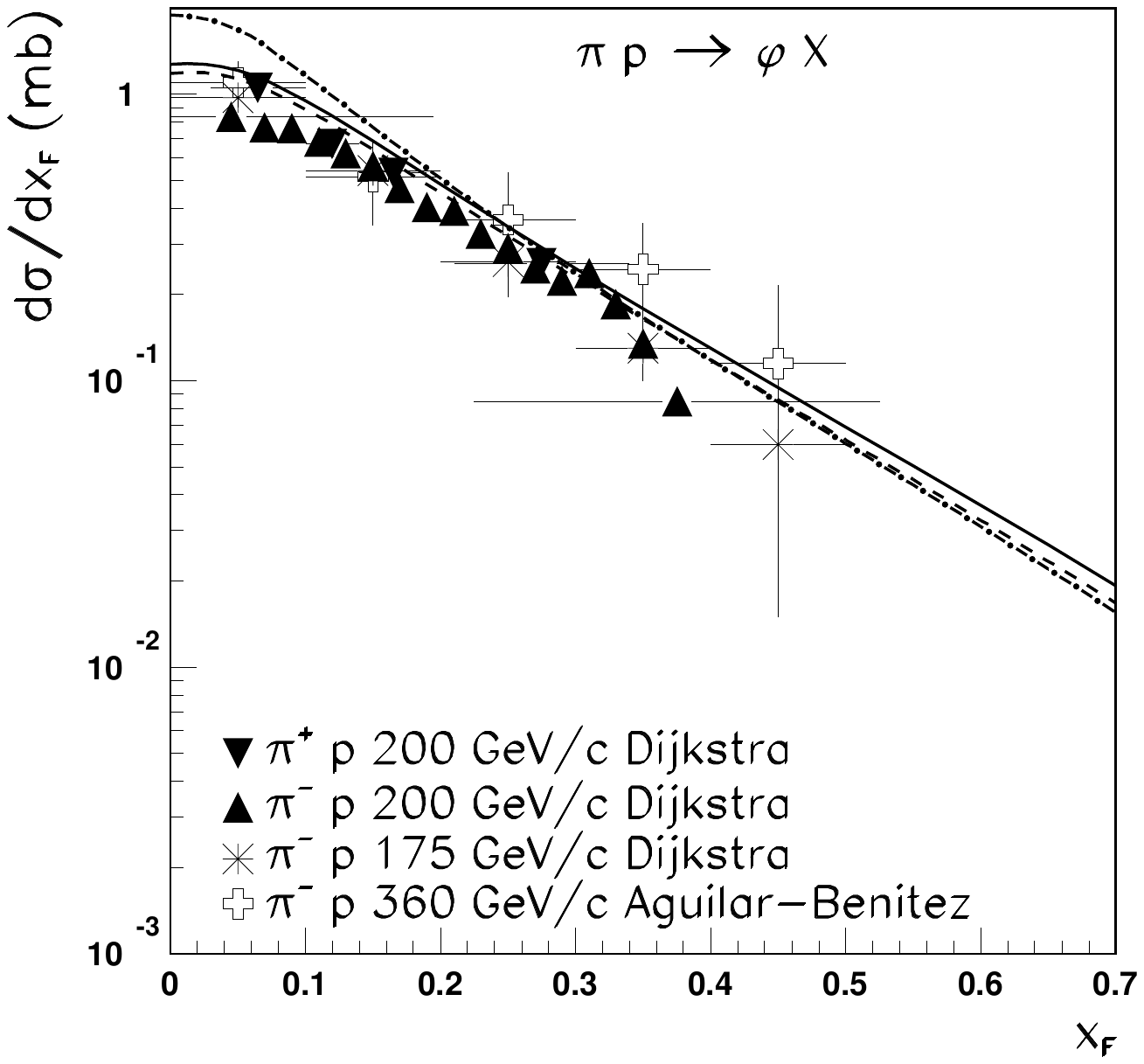}
\vskip -.7cm
\caption{\footnotesize
Two sets of experimental data on the $x_F$-spectra of $\varphi$-mesons produced 
in $\pi^{\pm} p$ collisions for two different initial momenta of $\pi$-mesons  93$-$140 GeV/c 
(upper panel) and 175$-$360 GeV/c (lower panel), measured by different collaborations 
\cite{apsimon,agbenpi,daum,dijkstra}, and the corresponding theoretical curves (see the main text).}
\end{figure}

\begin{figure}[htb]
\label{phi158}
\centering
\vskip -7.cm
\includegraphics[width=.85\hsize]{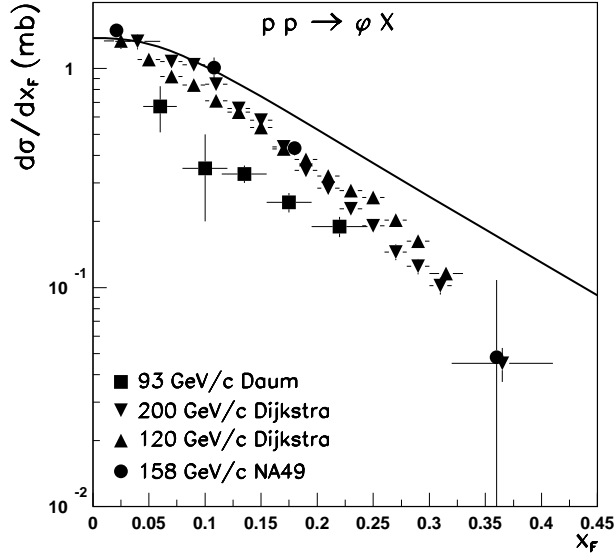}
\vskip -.7cm
\caption{\footnotesize
The experimental data on the $x_F$-spectra of $\varphi$-mesons produced in $pp$
collisions at different energies \cite{NA49,dijkstra,daum}, compared to the corresponding
QGSM calculation at 158 GeV/c (full curve).}
\end{figure}

\begin{figure}[htb]
\label{phi400}
\centering
\vskip -12.5cm
\includegraphics[width=.85\hsize]{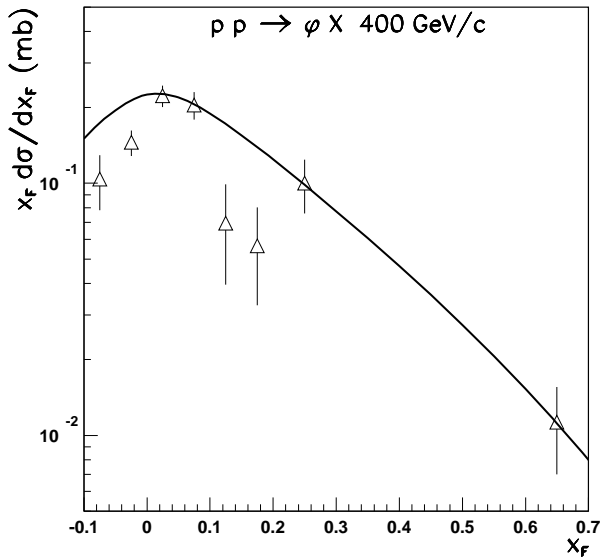}
\vskip -0.25cm
\caption{\footnotesize
The experimental data on the $x_F$-spectra $x_F d\sigma/dx_F$
of $\varphi$-mesons produced in $pp$ collisions
at 400 GeV/c~\cite{agben}, compared to the corresponding
QGSM calculation.}
\end{figure}

\begin{figure}[htb]
\label{phi158y}
\centering
\vskip -8.cm
\includegraphics[width=.9\hsize]{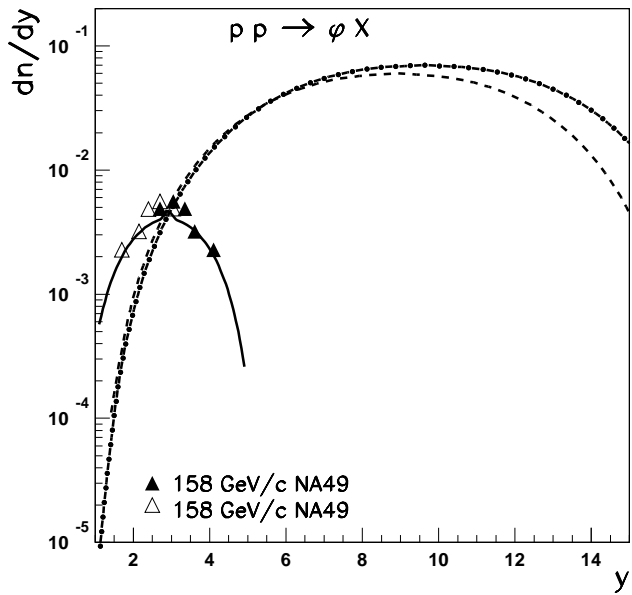}
\vskip -.8cm
\caption{\footnotesize
The experimental data~\cite{agben} on the $y$-spectra $dn/dy$ of $\varphi$-mesons
produced in $pp$ collisions at 158 GeV/c, and the corresponding QGSM result (full line).
The QGSM predictions for LHC energies 7 (dashed line) and 14 TeV( dashed-dotted line)
are also presented.}
\end{figure}

To calculate the $\varphi/\pi$ and $\varphi/K$ cross section ratios 
one needs the calculated values of both $\varphi$ and pseudoscalar $\pi$ and $K$-mesons
cross sections. For $\pi$ mesons the QGSM predictions are rather reliable~\cite{KaPi,Sh,MPS} 
up to the LHC energies, taking into account  in Eq.~(3) the growth of the $<p_T>$
of produced $\pi,K$ and $\varphi$- mesons with energy. 
\newpage

In Fig.~6 we present the QGSM description of the experimental data on
the inclusive spectra of $K^+$ (upper panel) and $K^-$ (lower panel) mesons
in $pp$-collisions on a wide energy range, going from 100 GeV/c up to 
$\sqrt{s}$=200 GeV~\cite{NA49,brenner}. 
The integrated over $p_T$ RHIC data at $\sqrt{s}$=200 GeV have been taken fron~\cite{NA49}.  
This data were obtained by the NA49 Collaboration in ref.~\cite{NA49} by 
interpolating the RHIC data at different $p_T$. 
 
In Fig.~6 one can appreciate that at low energies the normalization of 
the experimental data at 100 and 175 GeV/c in ref.~\cite{brenner} differs from
that of the experimental data in ref.~\cite{NA49} at 158 GeV/c.
We present the results of the QGSM calculations at two different energies: 
158 GeV/c ($\sqrt{s}$=17.3 GeV), by solid curves, and RHIC energy 
($\sqrt{s}$=200 Gev), by dashed curves.
One can see that the theoretical calculations at $\sqrt{s}$=200 Gev 
decrease alightly more rapidly than those at $\sqrt{s}$=17.3GeV.  

The energy dependence of the production cross section ratios of 
$\varphi/\pi^-$ (upper panel)~\cite{starphi,alvec} and $\phi/K^-$
(lower panel)~\cite{starphi,alfik} in pp collisions are 
presented in Fig.~7, where the QGSM description is shown by solid 
curves. The shapes of the two theoretical curves are similar because
the ratio of $K/\pi$ production depends rather weakly on the initial energy.
The discrepancies of the theoretical curves with the experimental points at
high energies in Fig.~7 can be connected to the 
differences in the kinematical windows for $\varphi$, $K$, and $\pi$ experimental
measurements at LHC energies.

\section{Conclusion}

The QGSM provides a reasonable description of Feynman $x_F$ and rapidity $y$ spectra of 
$\varphi$-meson production for the interaction of different hadron beams with a nucleon target in 
a wide energy region,  by only using one new normalization parameter, $a_{\varphi}$=0.11.
We show the QGSM prediction for $dn/dy$ cross sections for LHC energies.
We have also obtained a reasonable agreement for the $\varphi/\pi$ and $\varphi/K$ cross 
section ratios in a wide interval of the beam energy, going up to the LHC range.  

\begin{figure}[htb]
\centering
\vskip -6.cm
\includegraphics[width=.7\hsize]{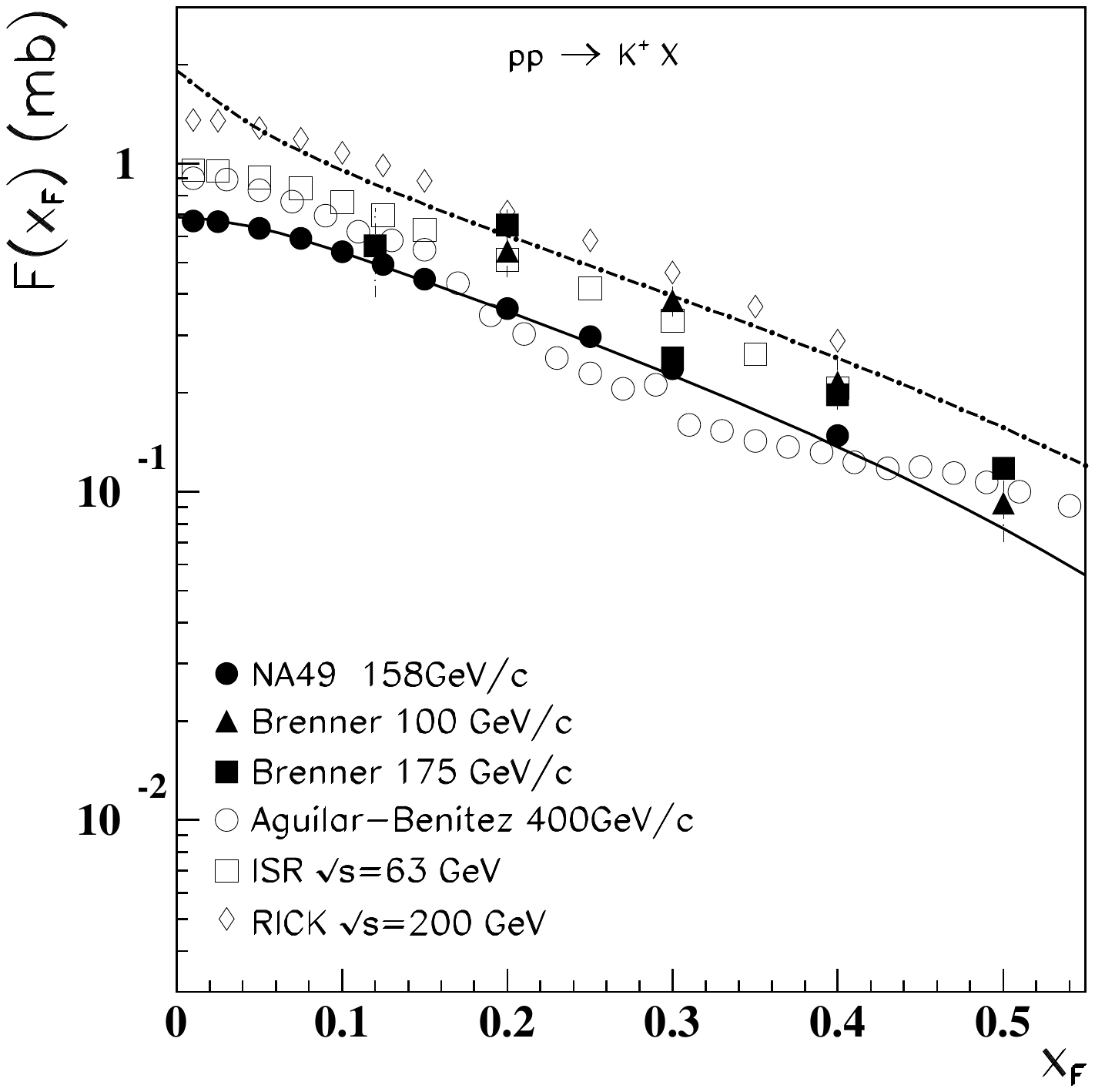}
\vskip -6.cm
\includegraphics[width=.7\hsize]{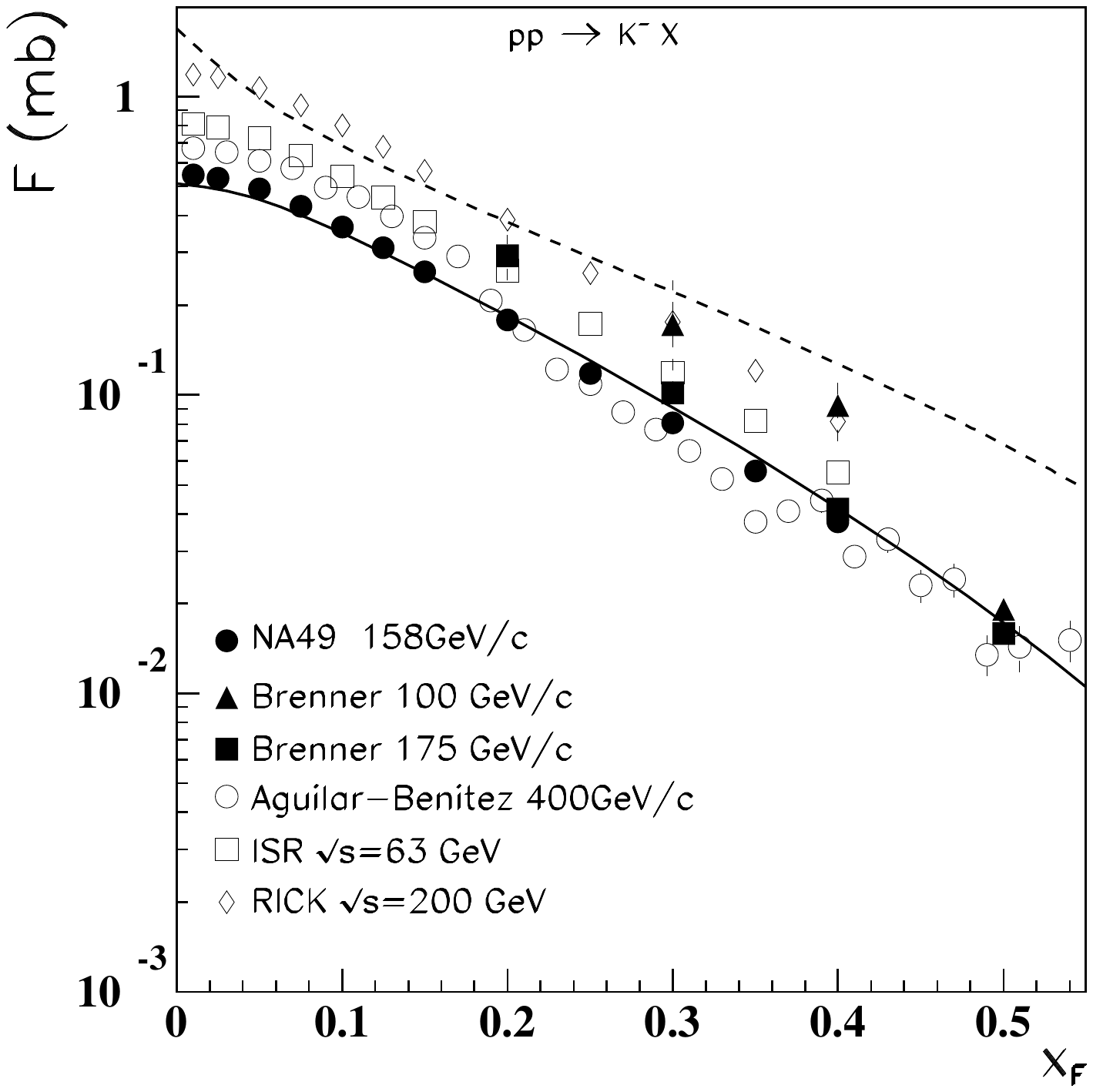}
\vskip -.75cm
\caption{\footnotesize
The QGSM description of the invariant cross section of $K^+$ (upper panel) and $K^-$ (lower panel) 
mesons produced in $pp$ collisions compared to the experimental data
at different energies~\cite{NA49,brenner}. Solid lines correspond to the QGSM result
at 158 GeV/c ($\sqrt{s}$=17.3 GeV), while dashed lines correspond to the QGSM calculation at RHIC energy 
($\sqrt{s}$=200 Gev).}
\end{figure}

\newpage
\begin{figure}[htb]
\centering
\label{ratios}
\vskip -6.5cm
\includegraphics[width=.7\hsize]{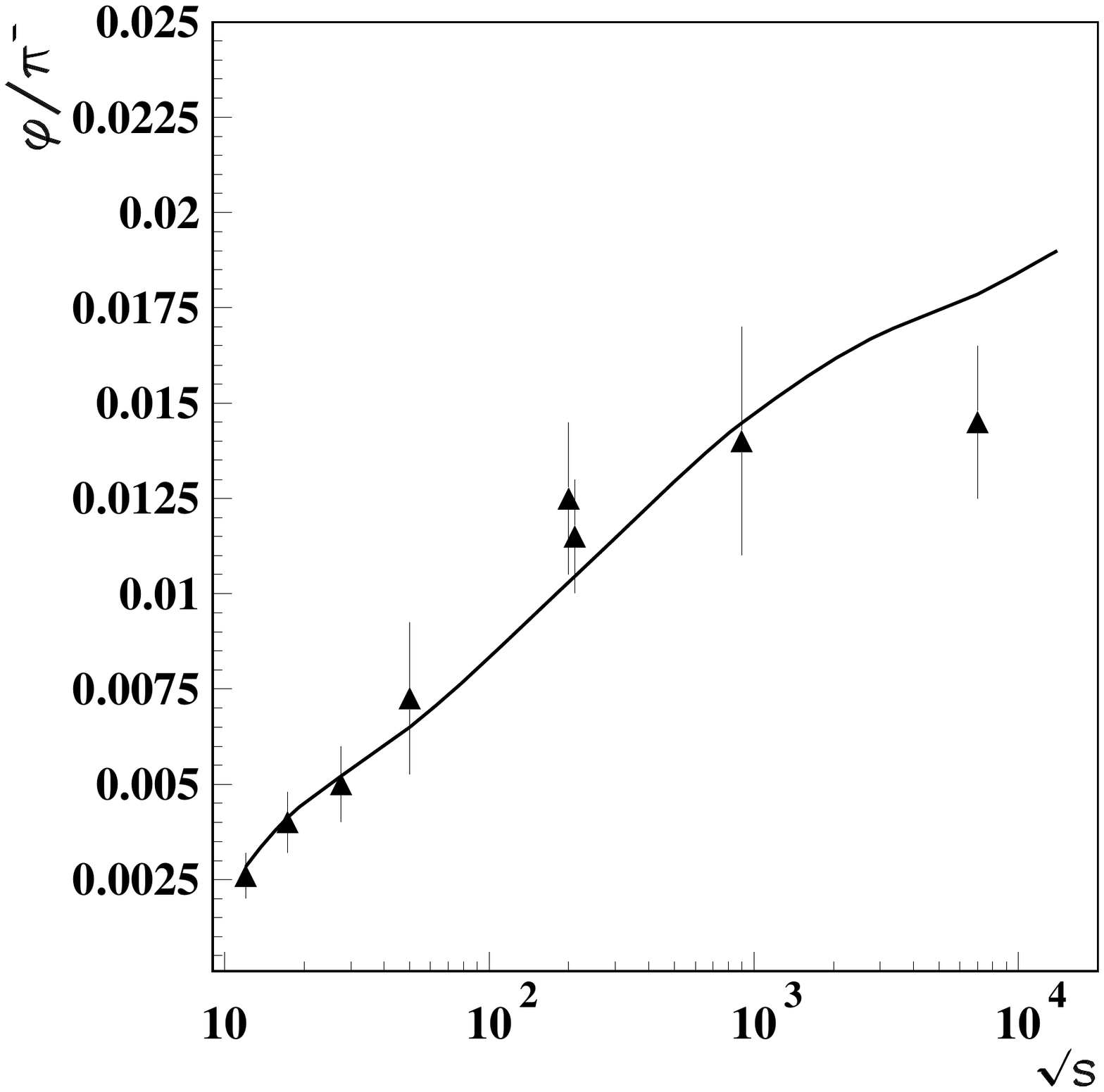}
\vskip -5.cm
\includegraphics[width=.7\hsize]{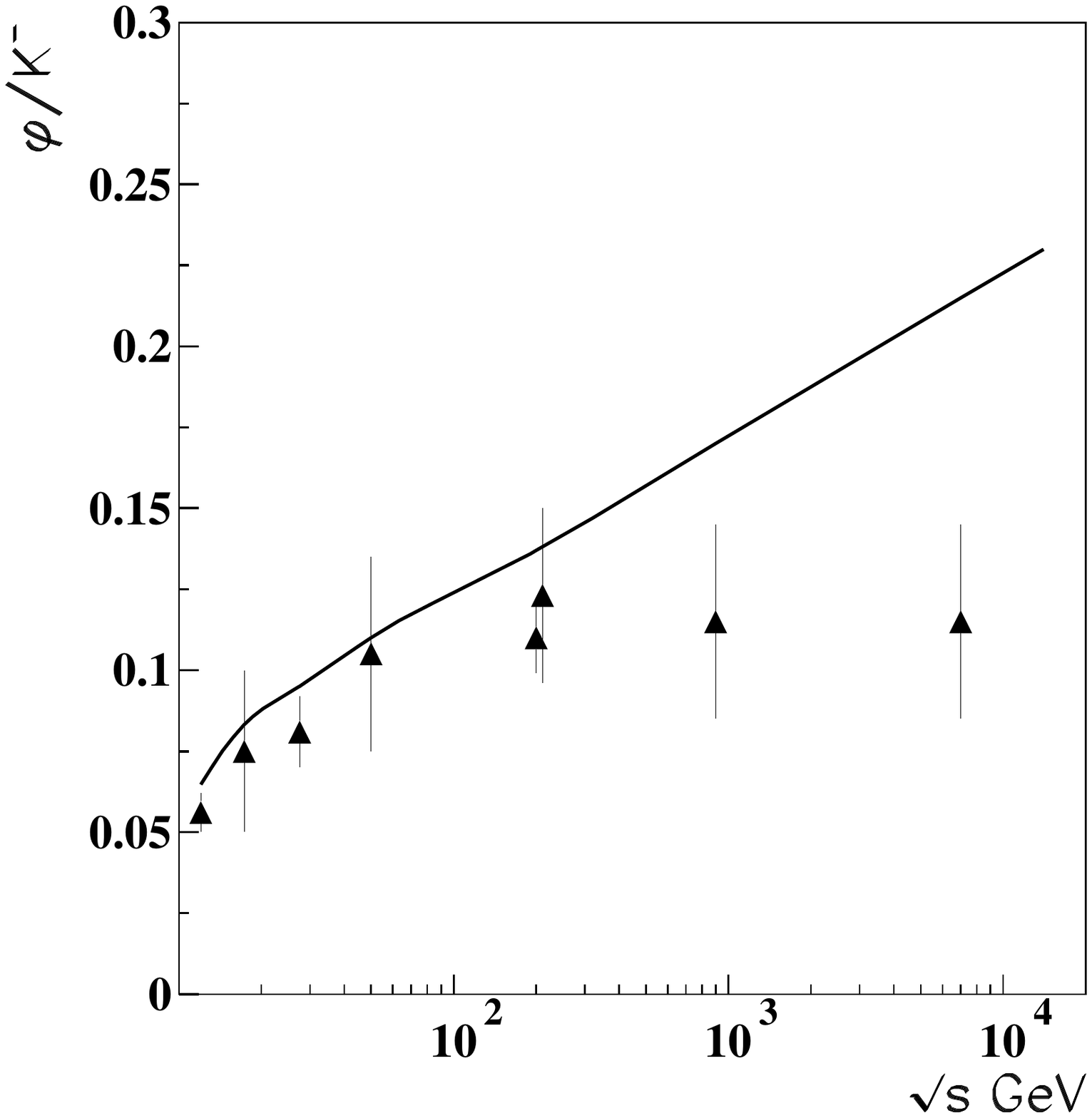}
\vskip -.5cm
\caption{\footnotesize
The QGSM description of the $\sqrt{s}$ dependence of $\varphi/\pi^-$ (upper panel)
and $\varphi/K^-$ (lower panel) cross section ratios produced in pp collisions,
compared to the corresponding experimental data in refs.~\cite{starphi,alvec,alfik}.}
\end{figure}

{\bf Acknowledgements}

We thank C. Pajares for useful discussions.
We are also grateful to N.I. Novikova for technical help.
This paper was supported by Ministerio de Econom\'i a y
Competitividad of Spain (FPA2011$-$22776), the Spanish
Consolider-Ingenio 2010 Programme CPAN (CSD2007-00042),
by Xunta de Galicia, Spain (2011/PC043), by the State
Committee of Science of the Republic of Armenia
(Grant-13-1C023), and, partially, by grant RSGSS-3628.2008.2.


\end{document}